\documentclass{article}
\usepackage{graphicx}
\usepackage{color}
\def\abstract#1{\vskip 7mm
        \begin{center}{\large Abstract}\par \smallskip
                \begin{minipage}[c]{12cm}
                        \small #1
                \end{minipage}
        \end{center}
}
\def\title#1{\begin{center}{\Large\bf #1}\end{center}}
\def\author#1{\vskip 5mm \begin{center}{#1}\end{center}}
\def\address#1{\begin{center}{\it #1}\end{center}}
\makeatletter
\@ifundefined{lesssim}{}{}
\@ifundefined{gtrsim}{}{}
\def\vereq#1#2{\lower3pt\vbox{\baselineskip1.5pt \lineskip1.5pt
\ialign{$\m@th#1\hfill##\hfil$\crcr#2\crcr\sim\crcr}}}
\makeatother

\begin{document}

\title{Hadron formation from interaction among quarks}
\author{Z. G. Tan \footnote{E-mail:tanzg@ccsu.cn}}
\address{$^1$Department of Electronic and Communication Engineering,Changsha University,Changsha, 410003,P.R.China\\
$^2$Key Laboratory of Quark and Lepton Physics (MOE) and Institute of Particle Physics, Central China Normal
University, Wuhan 430079,  China}
\author{C. B. Yang }
\address{$^2$Key Laboratory of Quark and Lepton Physics (MOE) and Institute of Particle Physics, Central China Normal
University, Wuhan 430079,  China}

\begin{abstract}
This paper deals with the hadronization process of quark system. A phenomenological potential is introduced
 to describe the interaction between a quark pair. The potential depends on the color charge of those quarks and their relative distances. Those quarks move according to classical equations of motion. Due to the color
 interaction, coloring quarks are separated to form color neutral clusters which are supposed to be the hadrons.
\end{abstract}

\section{Introduction}\label{sec1}
The study of particle production is an important subject in high-energy heavy-ion collisions. Physicists considered many mechanisms to describe the particle production processes,such as the string model \cite{string} and the
independent parton fragmentation model \cite{frag}. Both models can not explain the novel phenomena experimental
observed at BNL/RHIC, such as
an unexpectedly large $p/\pi$ ratio of about 1 at $p_T$ about 3 GeV/$c$ \cite{RHIC}, since they both predicted a very small $p/\pi$ ratio of about 0.2. In recent years, as a new approach to hadronization, the quark recombination model has been proposed \cite{ReCo}, which can be applied in any $p_T$ region, and can solve the puzzles from
 RHIC, such as the unexpectedly high $p/\pi$ ratio and the constituent quark number scaling of the elliptic flow.

 In the implementations of the quark recombination model as in \cite{ReCo}, it is assumed that the hadronization
 takes zero time, thus the quark distributions does not change in the process. Then some analytical expressions
 for the spectra of the final state particles can be derived. However, since the yield of meson (baryon) is
 proportional to the square (cubic) of the quark density, it will be four (eight) time larger if the quark density
 is doubled. So the naive quark recombination model violates the unitarity in the hadronization. To shun this
 difficulty, a finite hadronization time is introduced in Refs. \cite{fini}. It is assumed there that the
 production rate for a species of meson is proportional to the product of densities of corresponding quark
 and antiquark. Similar for baryon production rate. Thus, the essence of the original quark recombination
 model is kept in the revised model. Because of a finite time for hadroinization, production of particles
 will reduce the number of quarks and thus influence later particle production. Therefore, all particle productions
are correlated. In \cite{fini}, only the yields of particles have been studied, with the production correlation
fully considered. Frequently, one would like to learn the transverse spectra of produced particles. Such a job
cannot be done analytically, and some Monte Carlo method must be used.

   Another problem in the quark recombination model is about the interactions among quarks in quark-gluon-plasma
(QGP) produced in ultrarelativistic heavy ion collisions. The QGP created at BNL/RHIC and LHC/ALICE is not a weakly coupling  but strongly interacting matter. At the end of QGP evolution to the hadronization point, the interaction
among quarks may even be stronger. Then it is important to ask whether or not such interactions have influence
to the hadronization process.

This paper is a first step to solve the above two problems in the QGP hadronization. We will study hadron formation from a quark system with interactions by following the motion of quarks. We ask whether color neutral clusters can
be formed from color interactions when quarks move according to classical equation of motion. The organization of this paper is as follows. In Sec.\ref{sec2}, we will discuss potential for quark interactions,and give a possible form of it. Then in Sec.\ref{sec3} we present the numerical calculation details. Simulation results are shown in Sec.\ref{sec4}. The last section is for short discussions.

\section{Potential for quark interaction}\label{sec2}
\begin{figure}
\includegraphics[width=0.9\textwidth,height=0.60\textwidth]{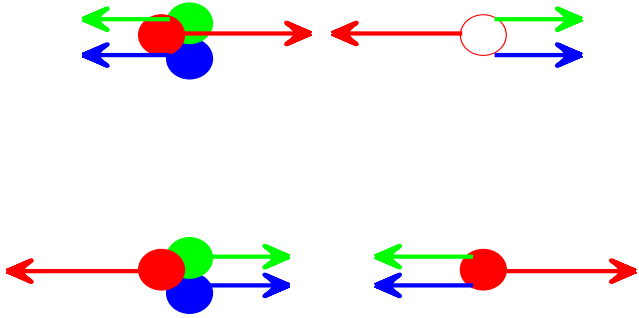}
\caption{According to Eq.\ref{eq5}, the total force that a baryon acts on a quark (lower) or a anti-quark (upper) is about zero.}
\label{f1}
\end{figure}

The interaction among quarks in hadronization should, in principle, be described by the basic theory, quantum
choromodynamics (QCD). When hadronization is concerned, as in the case discussed in this paper, the color
interactions among quarks can not be calculated perturbatively, and many non-perturbative effects play role since
hadronization process is a low momentum transfer process. Without first-principle guidance, one can use a
phenomenological potential to describe the quark interaction, as done for bound state problems. Such a potential
should depend on the colors carried by the interacting quarks. It has been shown that the potential corresponding
to a single-gluon exchange is inversely proportional to the separation $r$ between quarks when $r$ is small. When
$r$ is large, a string may be formed between two quarks and the corresponding potential is $\propto r$. Thus the
potential is assumed as
\begin{equation}
U_{ij}=c_{ij}a\left(r+\frac{b}{r}\right) \ ,\label{eq1}
\end{equation}
where $c_{ij}=c_{ji}$ is the color factor related quark $i$ and quark $j$, while $a$ and $b$ are two parameters of the model. If one chooses $a>0$, then some properties of $c_{ij}$ can be claimed. Since a quark and an anti-quark
with the same color  can combine to form a meson, they must attract one another, thus one may expect that
for such a quark pair $c_{ij}>0$. If a quark and an anti-quark do not carry the same color, $c_{ij}<0$. For two
quarks or two anti-quarks, a diquark can be formed if they carry different colors. Such a diquark is, in some
sense, similar to an anti-quark, in interacting with a third quark. Then for  two quarks or two anti-quarks,
$c_{ij}>0$ if they carry different colors. Otherwise, $c_{ij}<0$.
Now one can consider the interaction between a quark and a hadron. Since hadrons are color neutral, a quark would have no net force acting on a hadron if the hadron were a point particle. This condition can be satisfied if
one chooses $c_{r\bar{r}}=-c_{rr}=-2c_{rb}=-2c_{rg}$. Similarly for other color combinations.
Thus one can put all $c_{ij}$ as a matrix
\begin{equation}
C=Q\left[\begin{array}{rrr}-2 & 1&1\\1&-2&1\\1&1&-2\end{array}\right]\label{eq5}
\end{equation}
with $Q=-1$ for a quark and an anti-quark, and otherwise $Q=1$. The diagonal elements are for interaction
between quarks with the same color, and the off-diagonal ones for those interactions with different colors.
Then when two or three quarks move close enough to make a colorless cluster, the total force that act on an other quark is nearly zero, as shown in FIG.\ref{f1}.

The next step is to fix values of parameters $a$ and $b$ in Eq.(\ref{eq1}). While parameter $a$ determines the
strength of the total interaction between a pair of (anti) quarks, the parameter $b$ tells the relative strength
of the two interactions. Since the interactions among quarks are not considered in most event generators, one
may impose that the sum of the interaction potentials is equal to zero, in order to conserve the total energy of
the system. This additional requirement can be used to determine the value of parameter $b$. It can be understood
that the value of parameter $a$ will be responsible for the spatial separation between quarks in a hadron.
In this paper, we are only interested in whether color neutral clusters can be formed when color interactions are
taken into account. Thus the value of $a$ is not too relevant.

\section{Numerical calculation for quark system evolution}\label{sec3}
In order to see the feasibility of our model, we deal with the hadronization of  a quark system  with 30 quark and
antiquark pairs. It is assumed that the quarks have the same flavor. The generalization to include more flavors
is straightforward. The total energy is assumed to be E=60 GeV. We assume that they are taken from a large
thermal equilibrium distribution
\begin{eqnarray}
E&=&\sum_i\frac{gV}{\pi^2\hbar^3}\int f_i(T,p)p^0|\vec{p}|^2dp \nonumber\\
N&=&\sum_i\frac{gV}{\pi^2\hbar^3}\int f_i(T,p)|\vec{p}|^2dp, \label{therm}
\end{eqnarray}
where the summation runs over all flavors considered in the problem and $g=N_cN_s=6$ is the degeneracy number
for a quark. $f_i(T,p)$ is the distribution function for quarks of a specific flavor
\begin{equation}
f(T,p)=\frac{1}{1+e^{p^0/T}}\ ,\label{dis}
\end{equation}
with $p^0=\sqrt{|\vec{p}|^2+m_i^2}$. In the calculations, only $u, d$ and $s$ quark flavors are considered and
the corresponding masses are chosen as 0.3GeV for $u$ and $d$ quarks, and 0.5GeV for $s$ quarks, very close to
the one used in the constituent quark model. In fact, the main results obtained below have little dependence
on the number of flavors used.

For given $E$ and $N$, a initial volume $V$ and the temperature $T$ can be obtained from Eq.(\ref{therm}).
We put $N=60$ and $E=60$¡¡GeV in this paper. Then the value for $V$ is about 8.0
${\rm fm}^3$ and that for $T$ is 0.28GeV. We assume that the quarks are uniformly distributed in a spherical space with volume $V$. Direct sampling according to Eq. (\ref{therm}) for momentum distribution is not easy. But one can
do it in another way. One can assign to each quark a momentum with constant magnitude but
 a random direction to keep the total momentum zero. Then let the quarks undergo a certain number (for example 1000) elastic collisions inside the spatial volume $V$ with periodic boundary conditions used, as described
 in \cite{tan}. One can check that the momentum distribution obtained is like
 the one in Eq.(\ref{therm}).  Because of Eq. \ref{dis}, the momentum
 distribution is spherically symmetric.
 Now we add the color property to each quark randomly. For convenience we use 1,2,3 to represent the colors, say, 1 for red, 2 for green,and 3 for blue, and -1,-2,-3  for the correspondingly anti-colors. To make
 the whole system color neutral, one can simultaneously assign the opposite colors to quark and antiquark
 for each pair. Potential of the whole system can calculate according to
\begin{equation}
U=\sum_{i=1}^{N-1}\sum_{j=i+1}^N U_{ij}\ ,\label{eq6}
\end{equation}
where $U_{ij}$ is calculated according to Eq. (\ref{eq1}) with the fixed parameter $a=0.85$ GeV/fm$^{-1}$,
which is double that suggested in \cite{intro}. It should be mentioned that the value of parameter $a$
has no influence to the main results in this paper. The choice of this specific value for $a$ is just as a try.
$b$ could be determined by requiring $U=0$ at the the moment we assign colors to quarks. At any time $t$, we can calculate the force acted on quark i from quark j
\begin{equation}
\vec{f}_{ij}=-\bigtriangledown U_{ij}=c_{ij}a\left(\frac{\vec{r}_{ij}}{r_{ij}}-\frac{b\vec{r}_{ij}}{r^3_{ij}}\right)\ .\label{eq3}
\end{equation}
The total force acted on quark $i$ from all other quarks is
\begin{equation}
\vec{F}_{i}=\sum_{j\ne i}\vec{f}_{ij}\ .\label{eq4}
\end{equation}
After the system moves $dt$ forward,  particle $i$ will move according to the classical dynamics, and its
momentum and position are
\begin{eqnarray}
\vec{p}_i(t+dt)&=&\vec{p}_i(t)+\vec{F}_i dt\ ,\nonumber \\
\vec{r}_i(t+dt)&=&\vec{r}_i(t)+\vec{v}_i dt\ ,\label{eq9}
\end{eqnarray}
where $\vec{v}_i=\vec{p}_i/E_i$ and $E_i=\sqrt{|\vec{p}_i|^2+m_i^2}$.
The evolution of the system runs over a time period long enough to enable quark clusters well separated from
one another.

\section{Some results}\label{sec4}
\begin{figure}
\includegraphics[width=0.9\textwidth]{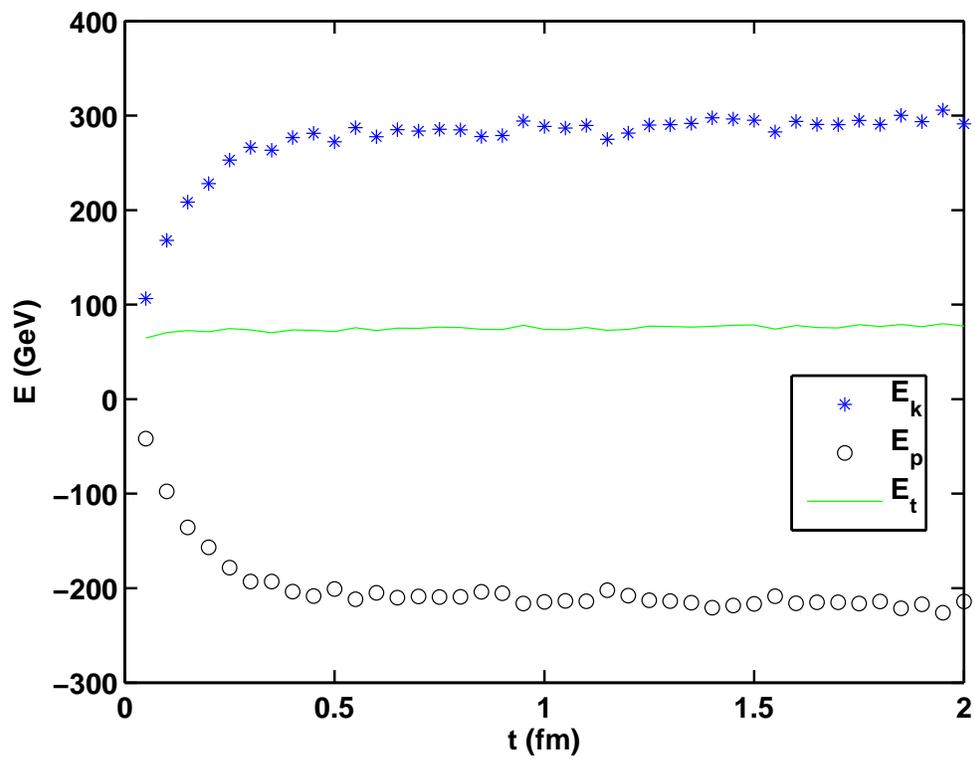}
\caption{Following the evolution of the system, the energy is transformed between the potential and kinetic. }
\label{f2}
\end{figure}

Since interaction among quarks depends strongly on positions of quarks, analytical calculation for all quarks'
trajectories is impossible. Thus numerical calculation has to be done with a finite time step $\Delta t>0$.
As a result of the finiteness of time step $\Delta t$, numerical errors must occur, and the total energy of
the system is, generally, not conserved exactly. In principle, only the $\Delta t\to dt\to 0$ the fluctuation
could be eliminated. In our calculation, we let $\Delta t=0.0001 $ fm/c, and assume that it is small enough.
The numerical results show that the energy is transformed between the potential and kinetic but the total energy is almost kept constant, as shown in FIG.\ref{f2}.

\begin{figure}
\includegraphics[width=0.9\textwidth]{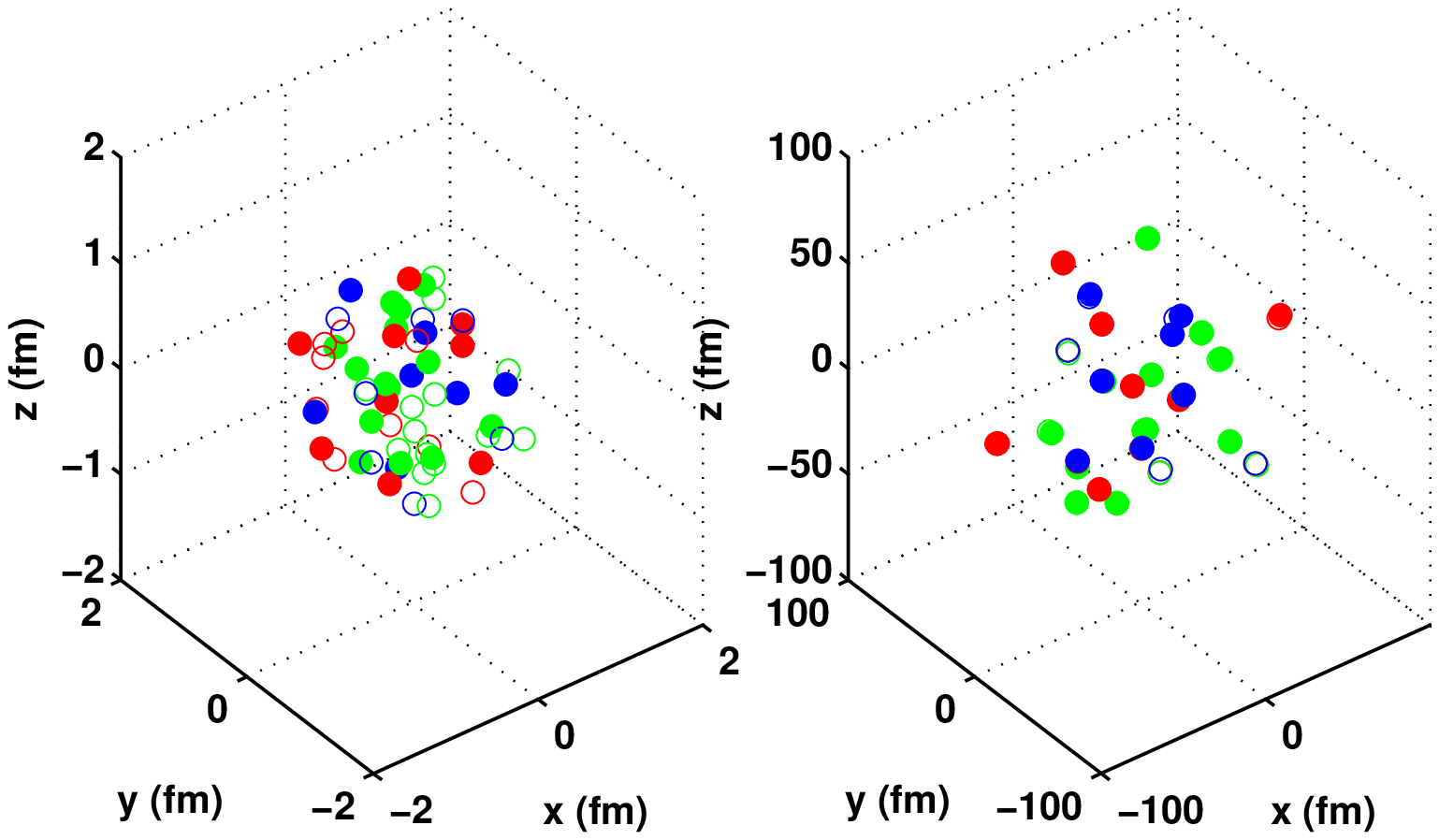}
\caption{Quark distributions for the initial and final state in the evolution of the system. Left part
is for the initial distribution of quarks and right one for that after a time of 100fermi/$c$.}
\label{f3}
\end{figure}

 With the method provided above, one can follow the evolution of our system. Actually, after a few dozens fermi/$c$
  both the potential and the kinetic energy of the system do not change anymore. That means that the system has
  been split into some clusters of color neutral and the interactions among those clusters is very weak.
  Thus one can say that
 hadrons or pre-hadrons has been formed then. In FIG.\ref{f3}, we show two instant pictures for the
 initial and final state distributions of the quarks in the space. From the final state picture many clusters with different size could be found clearly.

\begin{figure}
\includegraphics[width=0.9\textwidth]{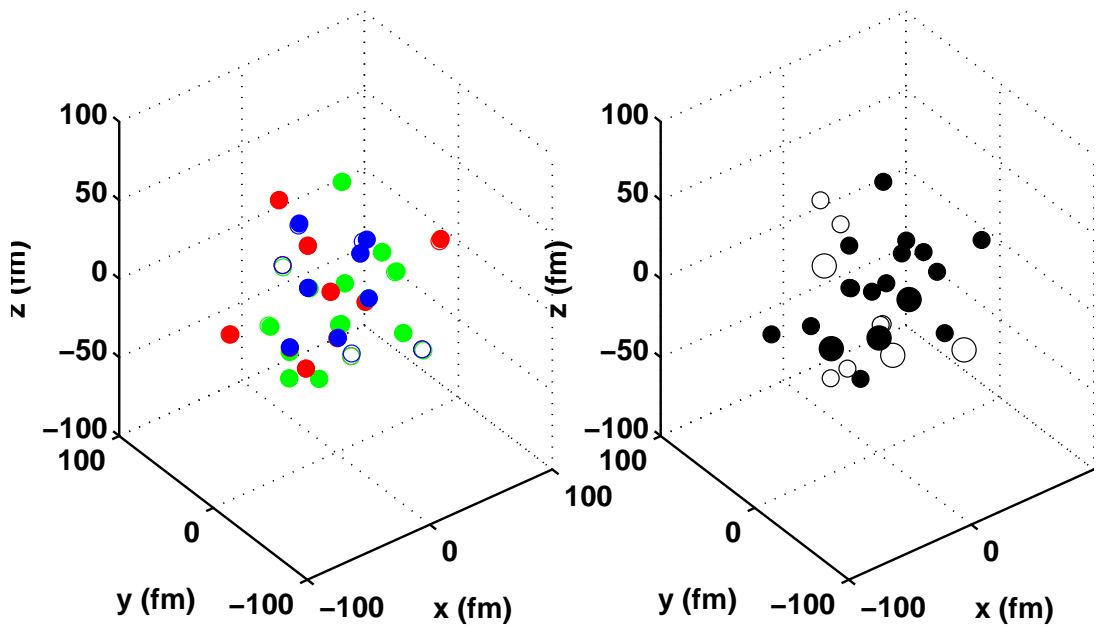}
\caption{Hadron spatial distribution obtained by replacing each cluster (in the left penal) by a hadron (in the right penal). Where a solid ball represents a particles, while a hollow one for an anti-particle. The size of the ball is used to distinguish meson and baryon.  A bigger symbol is for a baryon.}
\label{f4}
\end{figure}

Numerically, one can judge a cluster as a meson or a baryon or multi-quark state easily by counting the
quark numbers in the cluster. The species of the meson or baryon could be obtained by studying
the composition of each cluster. Comparing the distribution of particles in configuration space before and after hadronization as shown in FIG.\ref{f4}, we found that each cluster is exactly a hadron.

\section{Conclusion}
The hadronization of a quark system is considered as a dynamical process. An interaction potential is introduced
which depends on the colors and spatial separation of quarks. Quarks move according to classical equations of
motion. It is found that various hadrons could be formed naturally by gathering coloring quarks to color neutral clusters. In the process, the total energy is conserved within a high accuracy. Thus this paper shows that such
a method may be useful for studying more sophisticated hadronization processes in heavy ion collisions.

This work was supported in part by the National Natural Science Foundation of
China under Grant Nos. 11075061 and 11221504, by the Ministry of Education of China under Grant No.
306022 , and by the Programme of Introducing Talents of Discipline to Universities under Grant No.
 B08033,also by the Open innovation fund of the Ministry of Education of China under Grant No.
QLPL2014P01.

\end{document}